\documentclass[12pt,preprint]{aastex}
\shorttitle{Eta Carinae Event}
\shortauthors{Soker \& Hamann}

\def \cm{~\rm{cm}}

\def \K{~\rm{K}}
\def \g{~\rm{g}}

\def \AU{~\rm{AU}}
\def \erg{~\rm{erg}}

\def \yr{~\rm{yr}}

\def \astrobj#1{#1}

\begin{document}

\title{TRIGGERING ERUPTIVE MASS EJECTION IN LUMINOUS BLUE VARIABLES}

\author{Amos Harpaz and Noam Soker\altaffilmark{1}}

\altaffiltext{1}{Department of Physics, Technion$-$Israel
Institute of Technology, Haifa 32000 Israel;
phr89ah@physics.technion.ac.il; soker@physics.technion.ac.il.}

\begin{abstract}
We study the runaway mass loss process
of major eruptions of luminous blue variables (LBVs) stars, such
as the 1837-1856 Great Eruption of $\eta$ Carinae.
We follow the evolution of a massive star with a spherical stellar evolution
numerical code. After the star exhausted most of the hydrogen in the core
and had developed a large envelope, we remove mass at a rate of
$1 M_\odot \yr^{-1}$ from the outer envelope for 20 years.
We find that after removing a small amount of mass at a high rate,
the star contracts and releases a huge amount of gravitational energy.
We suggest that this energy can sustain the high mass loss rate.
The triggering of this runaway mass loss process might be a close stellar
companion or internal structural changes.
We show that a strong magnetic field region can be built in the radiative zone
above the convective core of the evolved massive star.
When this magnetic energy is released it might trigger a fast removal of mass,
and by that trigger an eruption.
Namely, LBV major eruptions might be triggered by magnetic activity cycles.
The prediction is that LBV stars that experience major eruptions should be
found to have a close companion and/or have signatures of strong magnetic activity
during or after the eruption.
\end{abstract}
\section{INTRODUCTION}
\label{intro}

Luminous Blue Variables (LBVs) are massive hot luminous stars.
They posses very strong winds that exhibit irregular variabilities on time
scales ranging from days to years.
On top of these variations, LBVs experience extreme mass loss rate episodes
(e.g., Smith \& Owocki 2006; Owocki \& van Marle 2009, and references therein), e.g.,
the 19th century eruptions of \astrobj{$\eta$ Carinae} (Humphreys et al. 1999), where a mass of
$\sim 10-20 M_\odot$ was lost (Smith et al. 2003b; Smith 2006; Smith \& Owocki 2006; Smith \& Ferland 2007).
These eruptions cannot be accounted for by the regular stellar luminosity,
and they require some extra energy source, e.g., internal structural change in the star,
that might even increase the stellar luminosity above the Eddington limit
(Owocki \& van Marle 2009).
However, part, or even all, of the increase in the luminosity of $\eta$ Car
in the 1837-1856 Great Eruption could have come from gravitational energy of
the mass accreted by the secondary star (Soker 2007).
The accretion of mass onto the secondary star can explain also the kinetic energy of
the Homunculus (Soker 2007); the Homunculus is the bipolar nebula of $\eta$ Car that
was formed in the Great Eruption (Davidson \& Humphreys 1997).

The influence of radiation on the mass loss process of stars near their Eddington luminosity
limit in relation to LBV eruptions is discussed by van Marle et al. (2008, 2009) and
Owocki \& van Marle (2009).
In particular, they discuss how the extended porosity formalism (Shaviv 1998, 2000)
can account for the Great Eruption of $\eta$ Car (Owocki et al. 2004;
van Marle et al. 2008, 2009).
In the present paper we do not deal with the interaction of radiation with matter.
We rather limit ourself to discuss the possible instability process that can lead to
the release of a huge amount of energy by internal structural change.

One of the significant differences between our approach and most other studies of the
Great Eruption of $\eta$ Car concerns the energy of the Homunculus.
While most studies (e.g., Smith 2006) attribute the entire energy source to the primary enhanced
luminosity, we take the view that most of the kinetic energy of the Homunculus results from
two opposite jets that were blown by the companion during the Great Eruption.
The companion blew the jets as it accreted mass from the primary dense wind via an
accretion disk (Soker 2007).
Therefore, we do not deal with the energy of the homunculus, but only with the energy that
is required to unbind a mass of $\sim 10-20 M_\odot$ during the Great Eruption
of $\eta$ Car, and similar LBV eruptions.
The same jets can account for fast ejecta that were blown from the $\eta$ Car binary system
(Smith \& Morse 2004).

The mass of $\sim 10-20 M_\odot$ (Smith et al. 2003b; Smith \& Ferland 2007) that was ejected in the
Great Eruption resides in the outer part of the radiative outer region of LBV stars.
In section 2 we build a stellar model that has a similar structure to that of
$\eta$ Car before the Great Eruption, and discuss some of its properties.
Soker (2007) already speculated that the Great Eruption of $\eta$ Car was triggered by
disturbances in the outer boundary of the inner convective region, most likely by
magnetic activity, that expelled the outer radiative zone.
Soker (2007) further mentioned that one way to form an extended envelope is by the
contraction of the inner layers.
In section 3 we go one step further and show that indeed, the
removal of the outer region of the star causes the star to shrink and
release a huge amount of gravitational energy.
Earlier suggestions for the cause of LBV instability was summarized by
Humphreys \& Davidson (1994). Some of them, e.g., the geyser model of
Maeder (1992), cannot work for a hot star like $\eta$ Car.
As we show, our model works for blue stars.
In section 4 we discuss in more detail the possibility that the initial mass
removal in LBV outbursts is triggered by magnetic activity.
We summarize in section 5.

\section{STELLAR STRUCTURE}
\label{stellar}

\begin{figure}  %
\begin{tabular}{cc}
{\includegraphics[scale=0.69]{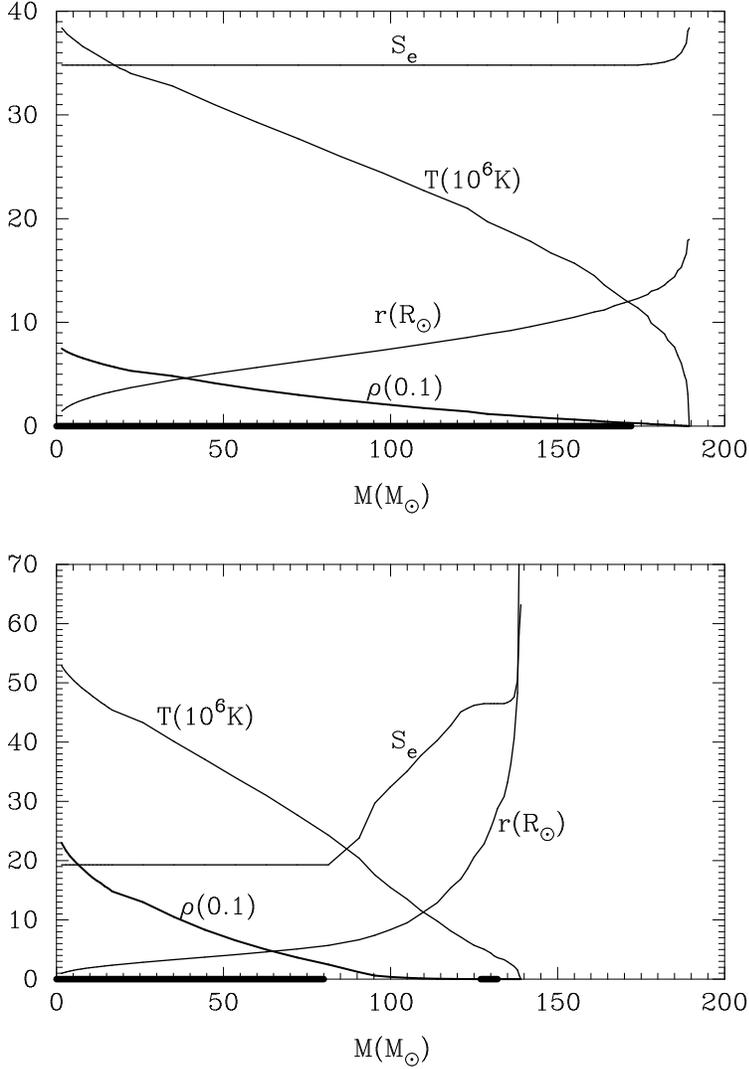}}
\end{tabular}
\caption{Density (in units of $0.1 \g \cm^{-3}$), temperature (in units of $10^6 \K$),
entropy $S_e$ (in relative units), and radius (in solar radii),
as function of mass for two evolutionary times: $t=0$ on the upper panel
and $t= 2.55 \times 10^6 \yr$ in the lower panel.
The two models have $[M(M_\odot),L(10^6L_\odot),T_e(10^4K)]=$
$(190,3,5.7)$, and $(139,3,1.6)$, respectively.
The thick lines on the horizontal axis mark the convective regions.
In the lower panel the photospheric radius is $220 R_\odot$,
and it is outside the graph.
}
\label{fig:models}
\end{figure}

We evolve a spherical stellar model with the same evolutionary code that
was used by us in previous studies over the years (for detail see Soker \& Harpaz 1999).
We start at $t=0$ with a zero-age main sequence star of mass $M_0=190 M_\odot$.
Mass loss is not a major part of our study as we are interested in the stellar model
toward the end of the main sequence. We simply set the mass loss
rate to be $\dot M = 2 \times 10^{-5} M_\odot \yr^{-1}$ (for more detail on the evolutionary track
of massive stars the reader can consult, e.g., Meynet \& Maeder 2003, 2005).
The mass, luminosity, and effective temperature, at 4 evolutionary points are
$[M(M_\odot),L(10^6L_\odot),T_e(10^4K)]=$
$(190,3,5.7)$, $(160,3,5.1)$, $(150,3,4.4)$, and $(139,3,1.6)$.
The luminosity does not evolve much, but as the hydrogen in the core is close
to exhaustion the envelope swells and the effective temperature decreases
(see also Smith \& Conti 2008).
In Fig. \ref{fig:models} we show the stellar structure at $t=0$
and at $t=2.55~$Myr.

Most relevant to us is the entropy profile.
The regions where the entropy profile is flat (actually decreasing very slowly)
are convective regions.
At early times the star is almost completely convective.
At later times the entropy is flat in the inner $\sim 80 M_\odot$.
The outer regions are mainly radiative. Above the inner convective region
the entropy increases substantially with mass (and radius).
Then, in the outer $\sim 15-20 M_\odot$ the profile becomes shallow,
and a second convective region exists there.
At late times most of the volume of the envelope is an outer extended region
with very low density ($\sim 10^{-7}-10^{-6} \g \cm^{-3}$) that contains a relatively small
amount of mass ($< 1 M_\odot$).

\begin{figure}  
\begin{tabular}{cc}
\hskip 2.0 cm
{\includegraphics[scale=0.99]{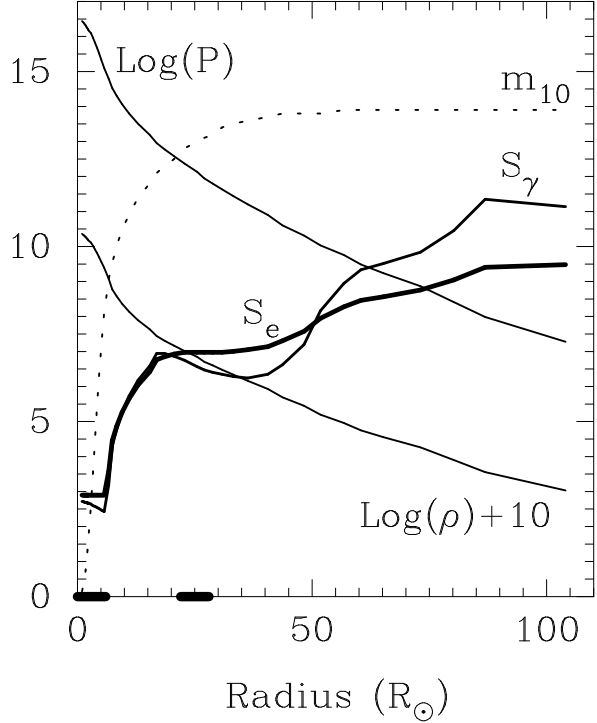}}
\end{tabular}
\caption{Logarithm of the density, log$\rho(g \cm^{-3})+10$, of the pressure, in c.g.s units,
mass in units of $10 M_\odot$, scaled accurate entropy from the stellar numerical code $S_e$, and
$S_\gamma \equiv 3.0\times 10^{-16} P \rho^{-4/3}$ in c.g.s. units, as function of radius
in solar radii.
There are two convective regions marked by thick lines on the horizontal axis.
One is in the core $r \la 6 R_\odot$ ($M \la 80 M_\odot$),
and the other at $22 \la r \la 28 R_\odot$ ($127 \la M \la 132 M_\odot$).
the photosphere is at a radius of $R_{\rm ph}=220 R_\odot$, but the region above $r=105 R_\odot$
contains very little mass, and is not shown in this figure.
The model is the one drawn in the lower panel of Fig. \ref{fig:models}:
$[M(M_\odot),L(10^6L_\odot),T_e(10^4K)]= (139,3,1.6)$.}
\label{fig:entropy}
\end{figure}

The evolutionary numerical code calculates the entropy $S_e$ using the full equation of state.
To further elaborate on the entropy behavior to be used later, we examine the quantity
$S_\gamma=P \rho^{\gamma_{\rm ad}}$. As evident from Fig. \ref{fig:entropy},
in the massive radiative region above the convective core a value of
$\gamma_{\rm ad}=1.33$ accurately describes the rapid entropy rise.
In Fig. \ref{fig:entropy} we plot the logarithm of the pressure and of the density, the mass,
the accurate entropy calculate by the stellar code $S_e$,
and of $S_\gamma$ for $\gamma_{\rm ad}=4/3$ (units are given in the caption),
as function of stellar radius for the second model shown in Fig. \ref{fig:models}.
We note again the rapid rise in the entropy from the core, and then the flattening in the
outer $\sim 15-20 M_\odot$, where a second convective region resides in the region
$22 \la r \la 28 R_\odot$. The very extended outer region is not shown, as it contains
a small amount of mass $< 1 M_\odot$.

The important property to take from the graphs is that the mass expelled in eruptions of LBVs,
such as the Great Eruption of $\eta$ Car, is a high-entropy gas.

\section{THE ERUPTION PHASE}
\label{massloss}

Stars with a radiative envelope shrink as they loss mass on a time scale shorter than
the thermal time scale (Webbink 1976; Heisler \& Alcock 1986; Maeder 1992).
The release of gravitational energy by the contracting envelope can lead to an
increase in the mass loss rate, resulting in a runaway mass loss process.

\begin{figure}  
\begin{tabular}{cc}
\hskip 2.0 cm
{\includegraphics[scale=0.99]{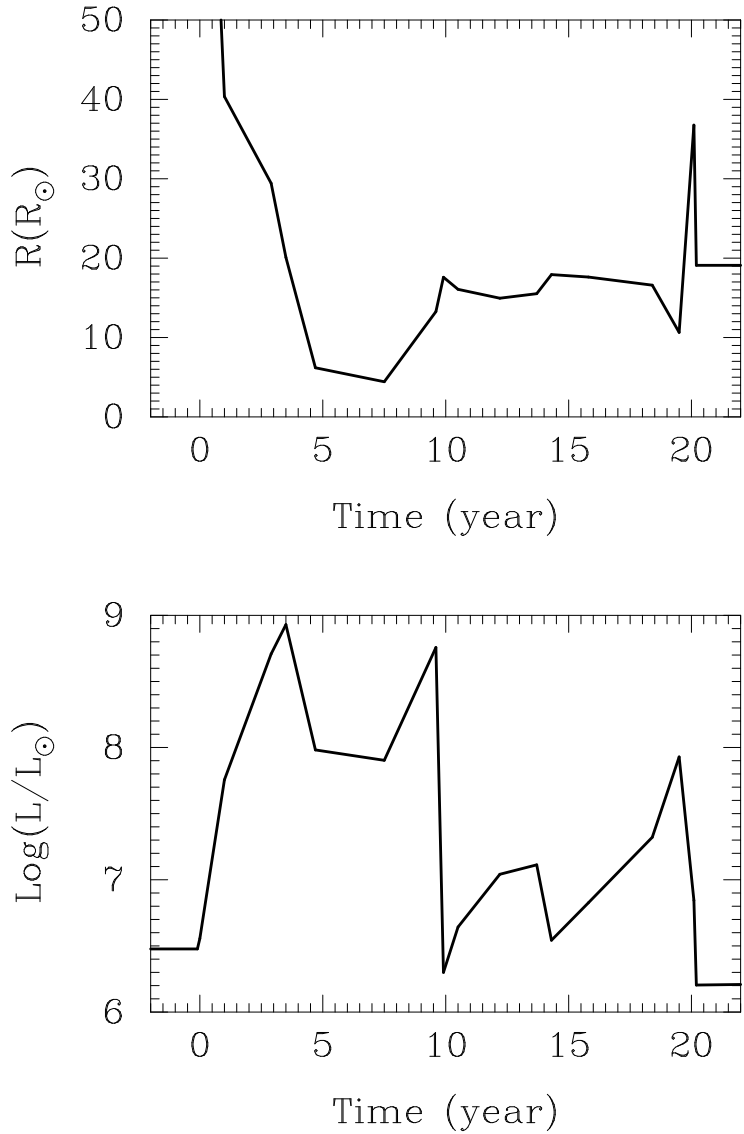}}
\end{tabular}
\caption{The radius and luminosity of the star during the runaway mass loss episode (eruption),
starting two years before the high mass loss rate and ending two years after.
The pre-outburst radius in our model is $\sim 200 R_\odot$.
During the eruption phase, $0 \le t_e \le 20 \yr$, mass was removed from the outer radiative
zone at a rate of $\dot M_{\rm burst} = 1 M_\odot \yr^{-1}$.
Because the mass loss time scale is shorter than the
thermal time scale, the numerical code is unstable, and there are large
fluctuations. However, the general shrinkage of the star and the release of a huge amount
of energy is evident. }
\label{fig:burst}
\end{figure}

To examine the behavior of our model we start with the star at the
evolutionary point $[M(M_\odot),L(10^6L_\odot),T_e(10^4K)]= (139,3,1.6)$,
and we remove $\Delta M_{\rm burst}=20 M_\odot$ with a constant mass loss rate
of $\dot M_{\rm burst} = 1 M_\odot \yr^{-1}$ for 20 years.
The mass is removed from the outer radiative region.
This mass loss rate mimics the average mass loss rate during the Great Eruption.
We then reduce the mass loss rate to $\dot M_p=2 \times 10^{-4} M_\odot \yr^{-1}$,
and follow the star for another 200 years.
In Fig. \ref{fig:burst} we plot the radius and luminosity of the star as function of time
during the eruption.

The mass loss rate during the 20 years eruption proceeds on a time scale much longer
than the dynamical time scale, but it is shorter than the thermal time scale.
At the beginning of the outburst the stellar radius is $R=220 R_\odot$.
The average thermal time scale of the outer region of mass $dm$ is
$\tau_{\rm th}=GMdm/RL$, while the mass loss time scale is $\tau_{\rm ml}= dm/\dot M$.
Their ratio is
\begin{equation}
\frac{\tau_{\rm th}}{\tau_{\rm ml}} = 7
\left( \frac{M}{140 M_\odot} \right)
\left( \frac{\dot M}{1 M_\odot \yr^{-1}} \right)
\left( \frac{R}{200 R_\odot} \right)^{-1}
\left( \frac{L}{3 \times 10^6 L_\odot} \right)^{-1} .
\label{eq:taus}
\end{equation}
The thermal time scale is substantially longer than the mass loss time scale.
As a result of this the star losses its thermal equilibrium and
rapidly contracts, i.e., on a time scale of few years which is much shorter than the
thermal time scale.
As our model is not fully built to take into account evolution on time scales shorter
than the thermal time scale, e.g., it does not take into account the energy required
to remove the mass from the envelope, our results of the rapid mass loss episode are
not accurate.
The code cannot handle properly the removal of mass during a time shorter than the
thermal time scale, and there are large fluctuations in the luminosity and radius.
In reality, the star exceeds its Eddington luminosity limit and is expected to be
unstable (Shaviv 2001).
However, as the mass loss time scale is longer than the
dynamical time scale, our results correctly give the general description.

It is evident that as mass loss starts on a timescale shorter than the thermal time scale
the inner regions of the star contracts, and release a huge amount of gravitational energy.
In our non-dynamical model the released gravitational energy leads to a higher luminosity.
In reality, the increased luminosity will result in a higher mass loss rate,
even if with a smaller terminal speed than in the regular wind, because the
luminosity exceeds the Eddington limit luminosity by more than an order of magnitude.
In particular, the opacity of the expelled mass is huge, and the extra luminosity would
be absorbed in the wind, and accelerates it. Our model does not have the wind component.
Therefore, the luminosity given above is not the observed luminosity, but rather the energy
available to accelerate the expelled mass.

The total kinetic energy of the gas in the Homunculus is
$\sim 5 \times 10^{49} (M_H/20 M_\odot) \erg$
where $M_H$ is the mass in the Homunculus (Smith 2006; Soker 2007).
The total gravitational energy liberated by the contracting star is $\sim 5 \times 10^{50} \erg$, which
is much larger. However, as mentioned, our numerical code does not have in
it the energy that is required to remove the mass.
After losing only $4 M_\odot$ the stellar radius has shrunk to $\sim 15 R_\odot$.
Losing another $16 M_\odot$ from an average radius of $15 R_\odot$
requires an energy of $\sim 5 \times 10^{50} \erg$. Namely, most of the energy will
go to unbind the expelled gas, and accelerate it.
As mentioned earlier, in the binary model (Soker 2001, 2007) the accreting companion can
supply most of the energy of the Homunculus, and energy considerations is not of a worry.

Our model is not fully consistent in another manner.
The numerical code does not include the triggering process of the high mass loss
rate during the eruption.
For example, the triggering can come from a tidal interaction with the companion
which deposited energy to the envelope.
Tidal enhancement of mass loss rate during the Great Eruption was mentioned by Soker (2001),
and during present periastron passages by Smith et al. (2003a).
Indeed, at the onset of the eruption the stellar radius was $\sim 1 AU$, while the
periastron distance in $\eta$ Car is $\sim 1.5 \AU$, implying that a strong tidal
interaction took place before and during the onset of the eruption.
The primary star could have gone through a weak instability (pulsation) that increased
its radius a little. If this continued while the secondary star was at periastron passage,
then the secondary could have remove some amount of mass, and trigger the outburst.
Another possible triggering process is discussed in the next section.

Although the photospheric radius decreases by more than an order of magnitude during
the eruption, mass shells do not contract much. The main contraction of the
photosphere is due to removal of the outer high entropy layers.
On average, during the 20 years eruption the mass shells at the photosphere
(each time a new mass shell) have their radii smaller by a factor of
$\zeta_r \le 2$ than their radii at the beginning of the eruption.
This implies that the ratio of rotation velocity to break-up velocity of
the mass shells increases by an average factor of only $\zeta_r^{1/2}  \simeq 1.5$.
In any case, even after contraction the mass shells at the photosphere have their
radii larger than their main sequence value.
This suggests that a single star cannot posses fast rotation.
This discussion further supports the claim that single star models cannot explain the bipolar
structure of the Homunculus, and an interaction with a companion is required (Soker 2004, 2007).

After we terminate the high mass removal rate, i.e., the end of the eruption phase,
and set the mass loss rate to $\dot M_p=2 \times 10^{-4} M_\odot \yr^{-1}$,
the star relaxes on a very short time, and starts a slow recovery.
While during the removal of mass at a high rate the stellar layers contract
on average and release gravitational energy, on the recovery phase the star
re-expands. This requires energy, and the luminosity is lower than its
pre-eruption value. It starts with $L=1.6 \times 10^6 L_\odot$
immediately after the eruption ends, and increases slowly to $L=2 \times 10^6 L_\odot$
at 200 years after the eruption.
This slow recovery might be related to the secular brightness of $\eta$ Car.
In our model the effective temperature 200 years after the eruption is
very high, $T_e \simeq 5 \times 10^4 \K$. However, our treatment of the eruption
is not accurate, and it is possible that the recovery phase is more rapid
than what we find. In any case, the dense wind of the primary in $\eta$ Car
blocks the UV radiation from its photosphere (Hillier et al. 2001, 2006).

\section{MAGNETIC ACTIVITY}
\label{magnetic}

In $\S$\ref{massloss} we found that an initial rapid mass loss will lead to a runaway process.
The initial rapid mass loss might be triggered by a tidal interaction with a companion,
or from internal changes in the star.
The proximity of the star to its Eddington luminosity limit most probably is also
a key property of the star.
Here we examine a possible triggering by magnetic fields (Soker 2007).

In the evolved star just prior to eruption, the inner $\sim 80 M_\odot$ region
is convective (the convective core), while most of the envelope is radiative.
It is quite possible that a dynamo will be operating in the
boundary of the convective core and the radiative region above it
(e.g., Charbonneau \& MacGregor 2001; MacGregor \& Cassinelli 2003).
This boundary is located where the entropy starts to rise, at
$(M,r)=(81 M_\odot, 6R_\odot)$ in the model presented in the lower
panel of Fig. \ref{fig:models} and in Fig. \ref{fig:entropy}.
As can be seen, above this radius the entropy rises sharply, until it flatten again
in the outer $\sim 20 M_\odot$. The mass of $\sim 20 M_\odot$ above the steep entropy
region, is about equal to the mass expelled in the Great Eruption of $\eta$ Car.
In our model this is not a coincidence.
In the sun, magnetic flux tubes that are formed in the lower boundary of the
convective region buoy outward to the surface. However, in the case studied here the
magnetic flux tubes need to buoy through the radiative region, where entropy
increases outward. We now examine this situation.

We note that a more thorough study is conducted by MacGregor \& Cassinelli (2003)
who considered also stellar rotation and heat transfer.
The aim of our simpler treatment is to emphasize the important role that
the steep entropy rise in the radiative zone might play in forming a
strong magnetic field region.

\begin{figure}  %
\begin{tabular}{cc}
{\includegraphics[scale=0.69]{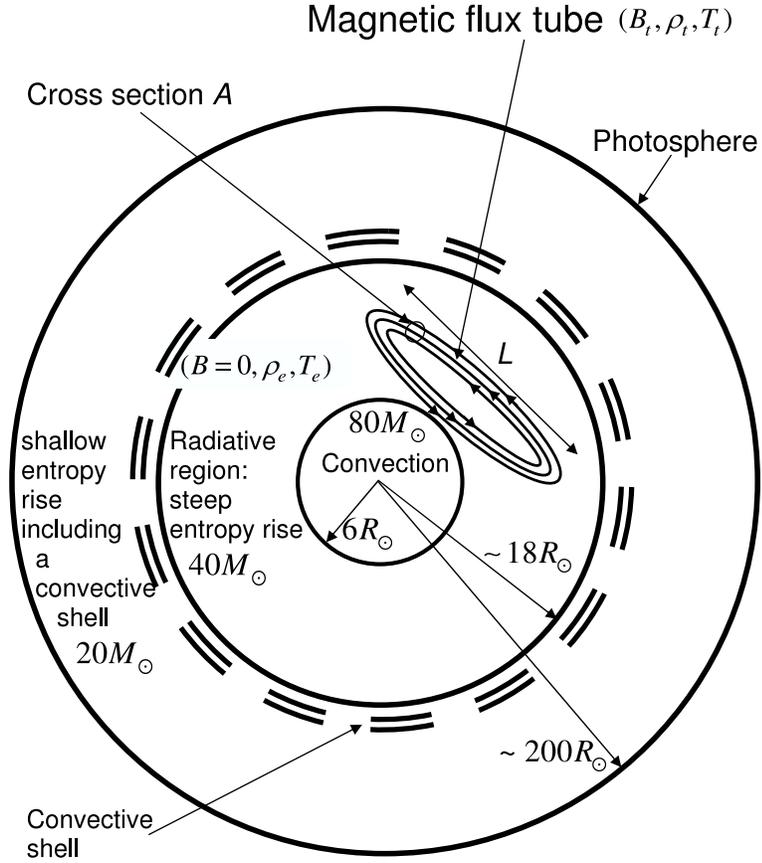}}
\end{tabular}
\caption{Schematic drawing of our stellar model just before the eruption,
with a flux loop embedded in the radiative envelope.
There is a core convective region in the sphere $r<r_0=6 R_\odot$,
and a second convective shell in the range $22 \la r \la 28 R_\odot$
($127 \la M \la 132 M_\odot$).}
\label{fig:flux}
\end{figure}

Consider a magnetic flux tube, or a magnetic flux loop, formed at the bottom of the
radiative zone $r=r_0$, with a magnetic field of intensity $B_{t0}$ and density
$\rho_{t0}$, and with an initial temperature equals to the ambient (envelope)
temperature $T_{t0}=T_e(r_0)\equiv T_{e0}$ at this radius.
The ambient density is $\rho_e$, and the magnetic field in the ambient medium is assumed
to be much weaker than that in the tube (or loop), such that the ambient pressure is
$P_e(r)=\rho_e k T_e/\mu m_H$, where $\mu m_H$ is the mean mass per particle.
Let the cross-section of the tube be $A(r)$, with $A_0 \equiv A(r_0)$, and let
its length be $L$.
The flux tube and our stellar model just before the eruption are drawn schematically in
Fig. \ref{fig:flux}.

For the evolution of the magnetic field inside the tube we take
\begin{equation}
B_t=B_{t0} \frac{A_0}{A}=B_{t0} \left( \frac{\rho_t}{\rho_{t0}} \right)^{\delta},
\label{eq:btube}
\end{equation}
where the first equality results from magnetic flux conservation, while the value
of $\delta$ depends on the relative variation of $A$ and $L$.
If the length of the tube does not change much as it rises, as might be the case if the
magnetic field is strong, then $\delta=1$.
If, on the other hand, the field is random or the relative expansion of all
dimensions is the same, as expected when the field is weak and it is dragged by the
rising tube, then $\delta=2/3$. We will take here $\delta=2/3$ as appropriate for a
weak initial magnetic field.

We also define the initial ratio of thermal to magnetic pressure inside the tube
\begin{equation}
\beta_0 \equiv \left( \frac {P_{\rm thermal}}{P_B} \right)_{t0}=\frac{\rho_{t0} k T_{t0}}{\mu m_H}
\left(  \frac{B_{t0}^2}{8 \pi}  \right)^{-1}.
\label{eq:beta}
\end{equation}
The pressure inside the tube includes magnetic and thermal contributions,
and can be written as
\begin{equation}
P_t={P_{\rm thermal}}+{P_B}=\frac{\rho_t k T_t}{\mu m_H}+ \frac{B_t^2}{8 \pi}.
\label{eq:ptube1}
\end{equation}
We assume adiabatic evolution
of the tube such that the temperature inside the tube
evolves as $T_t \propto \rho_t^{\gamma_{\rm ad}-1}$, and use
equation (\ref{eq:btube}) to write for the pressure inside the tube
\begin{equation}
P_t=\frac{\rho_{t0} k T_{t0}}{\mu m_H} \left( \frac{\rho_t}{\rho_{t0}} \right)^{\gamma_{\rm ad}} +
\frac{B_{t0}^2}{8 \pi} \left( \frac{\rho_t}{\rho_{t0}} \right)^{2 \delta}.
\label{eq:ptube2}
\end{equation}
The adiabatic index is not constant in the star. However, as evident from Fig. \ref{fig:entropy},
it is quite constant with $\gamma_{\rm ad}\simeq 4/3$ in the radiative zone above
the convective core (because radiation pressure is very important there).
In that zone, $6 \la r \la 22 R_\odot$ ($80 \la M \la 120 M_\odot$),
the entropy steeply rises by a factor of $S_e(22 R_\odot)/S_{e0}=2.3$ (Fig. \ref{fig:entropy}).

Using the definition of $\beta_0$ we can cast equation (\ref{eq:ptube2}) into
\begin{equation}
P_t=\frac{P_{e0}}{1+\beta_0^{-1}} \left[ \left( \frac{\rho_t}{\rho_{t0}} \right)^{\gamma_{\rm ad}} +
\frac{1}{\beta_0} \left( \frac{\rho_t}{\rho_{t0}} \right)^{2 \delta} \right],
\label{eq:ptube3}
\end{equation}
where $P_{e0}=P_{t0}=(1+\beta^{-1})\rho_{t0} k T_{t0}/\mu m_H$, as the tube and ambient
pressures at the origin $r_0$ are equal.
Since we assume that the temperature of the tube is equal to the ambient temperature
at its origin, we find the expression that connects the density of the tube to the
ambient pressure at the origin $\rho_{e0}=(1+\beta^{-1})\rho_{t0}$.
As the tube rises, its pressure equals to the ambient pressure $P_t=P_e(r)$.
Using these expressions for $\rho_{t0}$ and $P_t$, in equation (\ref{eq:ptube3}) gives
\begin{equation}
\frac{P_e(r)}{P_{e0}}=\frac{1}{1+\beta_0^{-1}}
 \left[ \left( \frac{\rho_t}{\rho_{e0}} \right)^{\gamma_{\rm ad}}(1+\beta_0^{-1})^{\gamma_{\rm ad}} +
\frac{1}{\beta_0} \left( \frac{\rho_t}{\rho_{e0}} \right)^{2 \delta}(1+\beta_0^{-1})^{2 \delta}
\right].
\label{eq:ptube4}
\end{equation}

We now examine the possibility that the flux tube reaches a new equilibrium position, namely,
a radius where its density is equal to the ambient (envelope) density $\rho_t=\rho_e(r)$.
This will be possible only if the entropy of the envelope increases fast enough with radius.
Substituting this equality in equation (\ref{eq:ptube4}), and dividing by
$[\rho_e(r)/\rho_{e0}]^{\gamma_{\rm ad}}$, we find
\begin{equation}
\frac{ P_e(r)/[\rho_e(r)]^{\gamma_{\rm ad}}}{ P_{e0}/\rho_{e0}^{\gamma_{\rm ad}}  }
=\left( {1+\beta_0^{-1}} \right)^{\gamma_{\rm ad}-1}
+
\left( {1+\beta_0^{-1}} \right)^{2\delta-1} \frac{1}{\beta_0}
 \left[ \frac{\rho_e(r)}{\rho_{e0}} \right]^{2 \delta -\gamma_{\rm ad}}.
\label{eq:equilibrium1}
\end{equation}
We take for the envelope entropy $S_e(r)\simeq S_\gamma(r) \equiv P_e / \rho_e^{\gamma_{\rm ad}}$,
with $\gamma_{\rm ad} \simeq4/3$, and for the behavior of the magnetic field
$2 \delta \simeq 4/3$, and derive from equation (\ref{eq:equilibrium1})
\begin{equation}
\frac{S_e(r)}{S_{e0}} \simeq \left( {1+\beta_0^{-1}} \right)^{1.33}
\label{eq:equilibrium2}
\end{equation}

Rising magnetic flux tubes will stop rising when the equality in
equation (\ref{eq:equilibrium2}) holds.
This situation is not completely stable.
MacGregor \& Cassinelli (2003) have showed that when heat transfer (by radiative diffusion)
and rotation are considered, the flux loops emitted in the equatorial plane can attain a
stationary equilibrium stable with respect to small displacements
in radius, but are unstable when perturbed in other directions.
However, the flux loops move outward on a long time scale (see below).
In any case, magnetic field energy can attain a large value in the radiative region
with the steep entropy rise.
The rise in entropy in the region
$6 \la r \la 22 R_\odot$ ($80 \la M \la 120 M_\odot$) is by a factor of
$S_e(22 R_\odot)/S_{e0}=2.3$ (Fig. \ref{fig:entropy}). From equation
(\ref{eq:equilibrium2}) we find then, that the magnetic flux tube will
come to rest in this radiative zone if $\beta_0 \ga 1$.
Namely, the initial magnetic pressure must be smaller than the thermal pressure in the
tube.

The rapid (on several dynamical time scales, or about a year) release of the magnetic energy
stored in the radiative region between the
two convective regions might trigger a high mass loss rate for a short time, that
will then sustain itself and develop to the runaway loss of $>10  M_\odot$ from
the envelope, as discussed in the previous section.

If stellar rotation and heat transfer are considered the flux tubes do not reach
an equilibrium (MacGregor \& Cassinelli  2003).
However, the rise of flux loops is determined by heating via radiative diffusion.
This heating proceeds on a long time scale, $>10^4 \yr$ (MacGregor \& Cassinelli  2003),
which is longer than the expected time between major LBV eruptions (Smith \& Owocki 2006).
Even meridional circulation (MacGregor \& Cassinelli  2003) will transport the
flux loops on a long time scale, much longer than buoyant time scale in a convective region.
In our scenario it is assumed that LBV stars posses magnetic activity cycles,
and the activity builds itself to a maximum on a time scale shorter than the meridional circulation
and heat transfer time scales.
The magnetic energy that is stored is required to remove a large amount of mass in a short
time to set the runaway mass loss episode.

In addition to the amplification of magnetic fields in the boundary of the
inner convective region and the radiative region above it, the outer convective
region can also amplify the magnetic field. All these possible dynamo
activities are beyond the scope of
the present paper.
Our only goal is to point to the possibility that LBV large eruptions are triggered by
magnetic activity.

\section{SUMMARY}
\label{summary}

Our main goal was to examine the triggering and runaway mass loss process
of major eruptions of luminous blue variables (LBVs) stars,
e.g., the 1837-1856 Great Eruption of $\eta$ Car.

We followed the evolution of a massive star.
We started with a solar composition and with
$[M(M_\odot),L(10^6L_\odot),T_e(10^4K)]=(190,3,5.7)$,
for the mass, luminosity and effective temperature, respectively.
We then evolved the star to almost complete depletion of hydrogen in
its core ($\S 2$), where the star had
$[M(M_\odot),L(10^6L_\odot),T_e(10^4K)]=(139,3,1.6)$;
our treatment of the atmosphere is not accurate, and for that the
effective temperature can be in the range $\sim 15,000-20,000 \K$.
The star has a convective core, then a radiative zone, followed by another convective shell
(Figs. \ref{fig:models} and \ref{fig:entropy}).
The outer most zone is a very extended radiative one, with very small amount of mass.
The most important property of the model is the development of a steep entropy rise
above the convective core.
The mass of $\sim 20 M_\odot$ above the steep entropy
region (see lower panel of Fig. \ref {fig:models} and Fig. \ref{fig:flux})
is about equal to the mass expelled in the Great Eruption of $\eta$ Car.
In our model this is not a coincidence.

In $\S 3$ we studied the response of the star to the eruption phase.
We removed $20 M_\odot$ at a rate of $1M_\odot \yr^{-1}$. Numerically,
the mass was removed from the outer radiative zone.
This mass loss time scale is shorter than the thermal time scale of the star
(eq. \ref{eq:taus}), and before removing even $1 M_\odot$, the envelope losses its thermal
equilibrium.
The envelope contracts, and its luminosity increases (Fig. \ref{fig:burst}).
The increase in luminosity that result from the release of gravitational energy
is huge. Very likely, this will sustain the high mass loss rate, and develop
into a runaway mass loss eruption.
Our code is not consistent in that we do not follow the energy required to
remove the mass; we simply remove mass from the envelope.
Still, we can safely conclude that if a triggering mechanism can remove mass,
even a small amount of $ \la 0.5M_\odot$, within a time much shorter than the
thermal time scale of that mass, a runway mass loss process will develop.

It should be noted that although the photosphere shrinks a lot, the motion of mass shells
inward is not by a large factor. This implies that the star will not spin-up much during the
eruption phase. The formation of a bipolar nebulae seems to require a binary companion.

In $\S 3$ we also followed the star for another 200 years after the end of the eruption,
with a mass loss rate of $\dot M=2 \times 10^{-4}$ to mimic the behavior of $\eta$ Car
after the Great Eruption.
As our treatment of the eruption phase is not fully consistent, we also
do not have the correct values for the post-eruption phase. For example, taking into
account the energy required to remove the mass from the star would reduce the energy radiated
by the star, and would make the envelope shrinkage smaller. The recovery phase
after the eruption will be shorter. Still, we managed to show that
the star tries to recover from the eruption.

The triggering of the rapid mass loss rate can come from a companion or from internal
structural changes. In the case of $\eta$ Car, a strong interaction with the companion
took place, and this could have triggered the eruption.
In $\S 4$ we examine the possibility that a magnetic activity cycle is the trigger.
We showed that a strong magnetic field region can be built in the radiative zone
above the convective core, as schematically drawn in Fig. \ref{fig:flux}.
When this energy is released as the magnetic cycle reaches its peak, it can
trigger a fast removal of mass, and by that trigger the eruption.

The prediction is that LBV stars that are experiencing major eruptions, should be
found to have a close companion, and/or have signatures of strong magnetic activity
during or after the eruption.

We thank Nathan Smith for useful comments.
This research was supported by grants from the Israel Science Foundation,
and from Asher Space Research Institute at the Technion.

\end{document}